\documentstyle[aps,prb]{revtex}

\begin{document}
\draft
\title{Enhancement of pair correlation in a one-dimensional hybridization model}
\author{Yupeng Wang$^{1,2}$ {~~}Jian-Hui Dai$^3${~~}Fu-Cho Pu$^{4,5}${~~}Ulrich Eckern$^1$ }

\address{
1.Institut f\"{u}r Physik, Universit\"at Augsburg, 86135 Augsburg,
Germany\\
2.Laboratory of Ultra-Low Temperature Physics, Chinese Academy of Sciences, P.O.
Box 2711, Beijing 100080, People's Republic 
of China\\
3.Condensed Matter Section, International Center for Theoretical Physics, P.O. Box 586, Treiste 34100, Italy\\
4.Department of Physics, Guangzhou Teacher College, Guangzhou 510400, People's Republic of China\\
5.Institute of Physics, Chinese Academy of Sciences, Beijing 100080, People's Republic of China}
\maketitle
\begin{abstract}
We propose an integrable model of one-dimensional (1D) interacting electrons coupled
with the local orbitals arrayed periodically in the chain. Since the local orbitals are introduced in a way that
double occupation is forbidden, the model keeps the main feature of the periodic Anderson model with an interacting
host. For the attractive
interaction, it is found that the local orbitals enhance the effective mass of
the Cooper-pair-like singlets and also the pair correlation in the ground state. However, the persistent current is 
depressed in this case.
For the repulsive interaction case, the Hamiltonian is non-Hermitian but allows Cooper pair
solutions with small momenta, which are induced by the hybridization between
the extended state and the local orbitals. 
\end{abstract}
\section{introduction}
Metallic compounds containing elements with partially filled $d$-shells or $f$-shells,
belong to the category of strongly correlated electrons. Two typical examples are
the high-Tc superconductors and the heavy fermion compounds in which the spin fluctuation may play 
central rule\cite{1,2}. The normal state properties of heavy fermion compounds are characterized
by a large Pauli susceptibility and specific heat as compared to conventional metals.
Such  phenomena are attributed to the large effective mass of the electrons near the
Fermi surface. These anomalous are generally believed to be due to the formation
of  resonant states at the Fermi level, which is induced by the admixture of local $f$-orbitals
and the conduction electrons, and therefore the systems are usually modeled by the periodic Anderson model
or the Kondo lattice model in some limiting cases. One of the major mysteries of
the heavy fermions is how superconductivity could be supported in a system with strong local moments\cite{2}.
It is generally accepted that magnetic impurities in BCS superconductors break the time reversal symmetry and are 
unfavorable to the formation of Cooper pairs. Such a pair breaking effect directly causes the reduction
of the energy gap of the superconducting state and the transition temperature\cite{3}. However,
the situations may be different in some strongly correlated electron systems, where the electrons
from the same source could be responsible for both the superconductivity and the magnetism\cite{1}. 
How magnetism and superconductivity reconcile each other is a hot interest in modern condensed matter physics
and still remains to be an open problem.
\par
Based on the development of the strongly correlated electron systems and low-dimensional systems, many
efforts have been done in recent years to understand how the magnetic impurities behave in a 1D correlated
host\cite{4}. Several integrable models were proposed\cite{5,6,7} to account for this problem and 
some novel features were found. In a recent paper\cite{8}, Schlottmann studied the attractive Hubbard model
with a finite concentration of magnetic impurities. He found the impurities generally weaken the binding
energy of the singlet pairs and the spin gap could be closed above a critical concentration of impurities.
\par
The quantum inverse scattering method (QISM) provides a powerful tool to construct
integrable models in 1D\cite{9}. In a lattice model, a local operator $L_{n,\tau}(\lambda)$ can be defined, which satisfy the following 
Yang-Baxter relation\cite{10}
\begin{eqnarray}
R_{\tau,\tau'}(\lambda,\mu)L_{n,\tau}(\lambda)L_{n,\tau'}(\mu)=L_{n,\tau'}(\mu)L_{n,\tau}(\lambda)R_{\tau,\tau'}(\lambda,\mu),
\end{eqnarray}
where $L_{n,\tau}(\lambda)$ acting on the auxiliary space $V_\tau$ and the quantum space $V_n$ respectively, $\lambda$ and
$\mu$ are the spectral parameters, $R_{\tau,\tau'}(\lambda,\mu)=L_{\tau,\tau'}(\lambda-\mu)$ is a c-number matrix. Define
the transition matrix $T_\tau(\lambda)$ as
\begin{eqnarray}
T_\tau(\lambda)=L_{1,\tau}(\lambda)\cdots L_{N,\tau}(\lambda),
\end{eqnarray}
where $N$ is the site number of the lattice. From (1) we can easily show that
\begin{eqnarray}
R_{\tau,\tau'}(\lambda-\mu)T_\tau(\lambda)T_{\tau'}(\mu)=T_{\tau'}(\mu)T_\tau(\lambda)R_{\tau,\tau'}(\lambda-\mu).
\end{eqnarray}
Tracing $\tau$ and $\tau'$ in the above equation we have
\begin{eqnarray}
[tr_\tau T_\tau(\lambda),tr_\tau T_\tau(\mu)]=0.
\end{eqnarray}
Suppose $\tau(\lambda)\equiv tr_\tau T_\tau(\lambda)$ allows the following expansion
\begin{eqnarray}
\tau(\lambda)=t_0+\frac{t_1}\lambda+\frac{t_2}{\lambda^2}+\cdots.
\end{eqnarray}
Since $\lambda$ and $\mu$ are arbitrary parameters, from (4) we obtain
\begin{eqnarray}
[t_m,t_n]=0, {~~~~~~~}m,n=0,1,2\cdots
\end{eqnarray}
Choosing one of $t_n$ as the Hamiltonian of the system, from (6) we know that all $t_m$ are conserved quantities. Therefore,
we can establish the common eigen states of these quantities. Generally, the generating operators of these eigen states are 
chosen from the off diagonal elements of the matrix $T_\tau(\lambda)$.
For the impurity models, the impurities are added
by including some inhomogeneous vertex operators in the transition matrix\cite{7}, which satisfy 
the same Yang-Baxter relation to that of the host. 
\par
In this paper, we consider a model of 1D interacting electrons coupled with the local orbitals arrayed periodically
in the chain. Maximumly, only one electron can occupy a single local state. Therefore, the model preserves the main feature of a 
1D periodic Anderson model in the limit $U\to\infty$. The structure of the present paper is the following: In the
subsequent section, the model Hamiltonian and its Bethe ansatz solution will be constructed based on QISM. In sect. III, 
we 
discuss the attractive interaction case. It is shown that the effective mass of the Cooper-pair-like singlets as well as the pair correlation
in the ground state are enhanced by the local orbitals. In sect. IV, we study the repulsive interaction case. Sec. V is attributed 
to the concluding remarks.
\section{the model and its Bethe ansatz}
In this paper, we construct an integrable model which describes interacting conduction
electrons in a continuum medium hybridizing with local orbitals. The atoms with local orbital are arrayed
periodically in the chain and the local states are described by the Hubbard operators
$X^n_{\alpha,\beta}\equiv |\alpha_n><\beta_n|$, ($\alpha_n,\beta_n=0,\uparrow,\downarrow$), with the constraint $X^n_{\uparrow\uparrow}
+X^n_{\downarrow\downarrow}+X_{00}^n=1$, which means double occupation of the same orbital is 
forbidden. Let us start with the following segment
transition matrix
\begin{eqnarray}
\partial_x T_{n+1}(\lambda|x,na)=:L(\lambda,x)T_{n+1}(\lambda|x,na):,
\end{eqnarray}
where $x\in [na,(n+1)a)$; $: :$ denotes the normal order of the fermions; $a$ is the space between two
nearest local orbitals; the $L$
operator is defined as
\[
L(\lambda,x)=\left(\begin{array}{ccc}
\frac i2\lambda &0 & i\sqrt{g}c_\uparrow(x) \cr
0&\frac i2\lambda&i\sqrt{g}c_\downarrow(x) \cr
-i\sqrt{g}c^\dagger_\uparrow(x) &-i\sqrt{g}c^\dagger_\downarrow(x)& -\frac i2\lambda
\end{array} \right),
\]
with $c_\sigma^\dagger(x)$ ($c_\sigma(x)$) the creation (annihilation) operator of the
conduction electrons. Here we take the boundary condition of $T_{n+1}(\lambda|x,na)$ as
\[
T_{n+1}(\lambda|na,na)\equiv L_n(\lambda)=\left(\begin{array}{ccc}
a'(\lambda)-b'(\lambda)X_{\uparrow,\uparrow}^n &-b'(\lambda)X_{\downarrow,\uparrow}^n& b'(\lambda)X_{0,\uparrow}^n\cr
-b'(\lambda)X_{\uparrow,\downarrow}^n &a'(\lambda)-b'(\lambda)X_{\downarrow,\downarrow}^n& b'(\lambda)X_{0,\downarrow}^n\cr
b'(\lambda)X_{0,\uparrow}^n &b'(\lambda)X_{0,\downarrow}^n& a'(\lambda)+b'(\lambda)X_{0,0}^n
\end{array}\right),
\]
where $a'(\lambda)=a(\lambda+ig/2)$, $b'(\lambda)=b(\lambda+ig/2)$ and
\begin{eqnarray}
a(\lambda)=\frac\lambda{\lambda-ig},{~~~~~}b(\lambda)=\frac{-ig}{\lambda-ig}.
\end{eqnarray}
We remark that with an unit boundary condition, the transition
matrix is just that of the $\delta$-potential Fermi gas model\cite{11} introduced by Yang\cite{10}. The non-unit boundary 
condition  is much similar to the inhomogeneous
$L$-operator in the lattice model\cite{7}. It is  easy to show that the following
Yang-Baxter relations hold
\begin{eqnarray}
R(\lambda-\mu)L(\lambda,x)\otimes_sL(\mu,x)
=L(\mu,x)\otimes_sL(\lambda,x)R(\lambda-\mu),\\
R(\lambda-\mu)L_n(\lambda)\otimes_sL_n(\mu)
=L_n(\mu)\otimes_sL_n(\lambda)R(\lambda-\mu),
\end{eqnarray}
where 
\begin{eqnarray}
R(\lambda)=a(\lambda)P+b(\lambda),
\end{eqnarray}
and $P$ is the FFB graded exchange operator acting on the direct product of the 
auxiliary spaces, $P^{a_1b_1}_{a_2b_2}=\delta_{a_1b_2}\delta_{a_2b_1}(-1)^{\epsilon_{b_1}\epsilon_{b_2}}$; 
$\epsilon_{\uparrow,\downarrow}=1, \epsilon_0=0$ (for convenience, we put the subscripts $\uparrow, \downarrow$ and $0$
as $1,2$ and $3$ respectively);  $\otimes_s$ denotes
the direct product with FFB grading\cite{12,11}
\begin{eqnarray}
(F\otimes_s G)^{ab}_{cd}=F_{ab}G_{cd}(-1)^{\epsilon_c(\epsilon_a+\epsilon_b)}.
\end{eqnarray}
From (9) and (10) we can easily derive
\begin{eqnarray}
R(\lambda-\mu)T_n(\lambda)\otimes_sT_n(\mu)=T_n(\mu)\otimes_sT_n(\lambda)R(\lambda-\mu),
\end{eqnarray}
with $T_n(\lambda)\equiv T(\lambda|na,na-a)$. Furthermore, the global transition matrix
\begin{eqnarray}
T(\lambda)\equiv T_N(\lambda)T_{N-1}(\lambda)\cdots T_1(\lambda)
\end{eqnarray}
satisfies the same Yang-Baxter equation of the segment ones $T_n(\lambda)$, 
\begin{eqnarray}
R(\lambda-\mu)T(\lambda)\otimes_sT(\mu)=T(\mu)\otimes_sT(\lambda)R(\lambda-\mu),
\end{eqnarray}
where
$N$ is the number of atoms with local orbital.
Introduce the notation
\[  
T(\lambda)=\left(\begin{array}{ccc}
A_{11}(\lambda) &A_{12}(\lambda)& B_1(\lambda)\cr
A_{21}(\lambda) &A_{22}(\lambda)& B_2(\lambda)\cr
C_1(\lambda) &C_2(\lambda)& D(\lambda)
\end{array} \right).
\]
From (15) we have the following commutation relations
\begin{eqnarray}
A_{ab}(\lambda)C_c(\mu)=(-1)^{\epsilon_a\epsilon_p}\frac{r(\lambda-\mu)^{dc}_{pb}}{a(\lambda-\mu)}
C_p(\mu)A_{ad}(\lambda)+\frac{b(\lambda-\mu)}{a(\lambda-\mu)}C_b(\lambda)A_{ac}(\mu),
\end{eqnarray}
\begin{eqnarray}
D(\lambda)C_c(\mu)=\frac1{a(\mu-\lambda)}C_c(\mu)D(\lambda)-\frac{b(\mu-\lambda)}{a(\mu-\lambda)}C_c(\lambda)
D(\mu),
\end{eqnarray}
\begin{eqnarray}
C_{a_1}(\lambda_1)C_{a_2}(\lambda_2)=r(\lambda_1-\lambda_2)^{b_1a_1}_{b_2a_1}C_{b_2}(\lambda_2)C_{b_1}(\lambda_1),
\end{eqnarray}
\begin{eqnarray}
[\tau(\lambda),\tau(\mu)]=0,
\end{eqnarray}
where
\begin{eqnarray}
\tau(\lambda)=str T(\lambda)=-A_{11}(\lambda)-A_{22}(\lambda)+D(\lambda)
\end{eqnarray}
and
\begin{eqnarray}
r(\lambda)^{ab}_{cd}=-b(\lambda)\delta_{ab}\delta_{cd}-a(\lambda)\delta_{ad}\delta_{bc}.
\end{eqnarray}
It  can be shown that $r(\lambda)$ satisfies the following Yang-Baxter relation
\begin{eqnarray}
r(\lambda-\mu)^{a_2c_2}_{a_3c_3}r(\lambda)^{a_1b_1}_{c_2d_2}r(\mu)^{d_2b_2}_{c_3b_3}=r(\mu)^{a_1c_1}_{a_2c_2}
r(\lambda)^{c_2d_2}_{a_3b_3}r(\lambda-\mu)^{c_1b_1}_{d_2b_2}.
\end{eqnarray}
From (19) we can see $\tau(\lambda)$ can be treated as a generator of an infinite number of conserved quantities.
Choose the vacuum state as
\begin{eqnarray}
c_\sigma(x)|0>=X_{0,\sigma}^n|0>=0.
\end{eqnarray}
We have
\[
T(\lambda)|0>=\left( \begin{array}{ccc}
e^{i\frac\lambda2L}a^N(\lambda+i\frac g2) &0& 0\cr
0&e^{i\frac\lambda2L}a^N(\lambda+i\frac g2)& 0\cr
C_1(\lambda)&C_2(\lambda)& e^{-i\frac\lambda2L}
\end{array}\right)|0>.
\]
Therefore, $C_a(\lambda)$ can be treated as the creation operators of the eigenstates of $\tau(\lambda)$:
\begin{eqnarray}
|k_1,\cdots,k_n|F>=C_{a_1}(k_1)C_{a_2}(k_2)\cdots C_{a_n}(k_n)|0>F^{a_n
\cdots a_1},
\end{eqnarray}
where the indices $a_j$ run over the values $1,2$ and $F^{a_n\cdots a_1}$ is a function of the spectral parameters
$k_j$. From the commutation relations (16) and  (17) we have
\begin{eqnarray}
D(\lambda)|k_1,\cdots,k_n|F>=e^{-i\frac\lambda2L}\prod_{j=1}^n\frac1{a(k_j-\lambda)}
|k_1,\cdots,k_n|F>\nonumber\\
+\sum_{l=1}^n({\bar \Lambda}_l)^{b_1\cdots b_n}_{a_1\cdots a_n}C_{b_l}(\lambda)
\prod_{j\neq l}^nC_{b_j}(k_j)|0>F^{a_n\cdots a_1},
\end{eqnarray}
\begin{eqnarray}
[A_{11}(\lambda)+A_{22}(\lambda)]|k_1,\cdots,k_n|F>=-e^{i\frac\lambda2L}a^N(\lambda+i\frac g2)
\prod_{j=1}^n\frac1{a(\lambda-k_j)}
\prod_{l=1}^n C_{b_l}(\lambda_l)|0>\nonumber\\
\times[\tau^{(1)}(\lambda)]^{b_1\cdots b_n}_{a_1\cdots a_n}F^{a_n\cdots a_1}
+\sum_{l=1}^n(\Lambda_l)^{b_1\cdots b_n}_{a_1\cdots a_n}C_{b_l}(\lambda)\prod_{j\neq l}^nC_{b_j}(k_j)|0>F^{a_n\cdots a_1},
\end{eqnarray}
where
\begin{eqnarray}
\tau^{(1)}(\lambda)=str[T_n^{(1)}(\lambda)]=str[L_n^{(1)}(\lambda-k_n)\cdots 
L_1^{(1)}(\lambda-k_1)],
\end{eqnarray}
and
\begin{eqnarray}
L_k^{(1)}(\lambda)=b(\lambda)P^{(1)}+a(\lambda)
\end{eqnarray}
which fulfill the Yang-Baxter relation
\begin{eqnarray}
r(\lambda-\mu)L_k^{(1)}(\lambda)\otimes_s L_k^{(1)}(\mu)=L_k^{(1)}(\mu)\otimes_sL_k^{(1)}(\lambda)r(\lambda-\mu),
\end{eqnarray}
with $(P^{(1)})^{ab}_{cd}=-\delta_{ad}\delta_{bc}$ the $4\times4$ permutation matrix;
\begin{eqnarray}
({\Lambda}_lF)^{b_1\cdots b_n}=\frac{b(\lambda-k_l)}{a(\lambda-k_l)}
\prod_{i\neq l}^n\frac1{a(k_l-k_i)}
a^N(k_l+i\frac g2)e^{i\frac{k_l}2L}F^{a_n\cdots a_lb_{l-1}\cdots b_1}\nonumber\\
\times (-1)^{l+1}L_n^{(1)}(k_l-k_n)^{b_ld_{n-1}}_{b_na_n}
 L_{n-1}^{(1)}(k_l-k_{n-1})^{d_{n-2}d_{n-3}}_{b_{n-1}a_{n-1}}\cdots 
 L_{l+1}(k_l-k_{l+1})^{d_la_l}_{b_{l+1}a_{l+1}},
\end{eqnarray}
\begin{eqnarray}
({\bar \Lambda}_lF)^{b_1\cdots b_n}=S(k_l)^{b_1\cdots b_l}_{a_1\cdots a_l}F^{b_n\cdots b_{l+1}a_l\cdots a_1}
[-\frac{b(k_l-\lambda)}{a(k_l-\lambda)}]e^{-i\frac{k_l}2L}\prod_{i\neq l}^n\frac1{a(k_i-k_l)},
\end{eqnarray}
where
\begin{eqnarray}
S(k_l)^{b_1\cdots b_l}_{a_1\cdots a_l}=r(k_{l-1}-k_l)^{b_{l-1}a_l}_{c_{l-1}a_{l-1}}\cdots r(k_1-
k_l)^{b_1c_2}_{b_la_1}.
\end{eqnarray}
To obtain the Bethe ansatz equations and the eigen values of $\tau(\lambda)$, we solve first the following eigen value problem
\begin{eqnarray}
\tau^{(1)}(\lambda)F=e(\lambda)F.
\end{eqnarray}
From (29) we know
\begin{eqnarray}
r(\lambda-\mu)T_n^{(1)}(\lambda)\otimes_sT_n^{(1)}(\mu)=T_n^{(1)}(\mu)\otimes_s T_n^{(1)}(\lambda)r(\lambda-\mu).
\end{eqnarray}
Introduce the notation
\begin{eqnarray*}
T_n^{(1)}(\lambda)=\pmatrix
{A^{(1)}(\lambda)&B^{(1)}(\lambda)\cr
C^{(1)}(\lambda)&D^{(1)}(\lambda)}.
\end{eqnarray*}
(34) gives
\begin{eqnarray}
D^{(1)}(\lambda)C^{(1)}(\mu)=\frac 1{a(\lambda-\mu)}C^{(1)}(\mu)D^{(1)}(\lambda)+\frac{b(\mu-\lambda)}{a(\mu-\lambda)}
C^{(1)}(\lambda)D^{(1)}(\mu),
\end{eqnarray}
\begin{eqnarray}
A^{(1)}(\lambda)C^{(1)}(\mu)=\frac1{a(\mu-\lambda)}C^{(1)}(\mu)A^{(1)}(\lambda)+\frac{b(\lambda-\mu)}{a(\lambda-\mu)}C^{(1)}
(\lambda)A^{(1)}(\mu),
\end{eqnarray}
\begin{eqnarray}
C^{(1)}(\lambda)C^{(1)}(\mu)=C^{(1)}(\mu)C^{(1)}(\lambda),{~~~~~~}[\tau^{(1)}(\lambda)(\lambda),\tau^{(1)}(\mu)]=0.
\end{eqnarray}
Define the pseudo vacuum $|0>^{(1)}$ as $B^{(1)}(\lambda)|0>^{(1)}=0$. The eigenstates
of $\tau^{(1)}(\lambda)$ can be written as
\begin{eqnarray}
|\mu_1,\cdots,\mu_M>=\prod_{\alpha=1}^MC^{(1)}(\mu_\alpha)|0>^{(1)}.
\end{eqnarray}
Acting $\tau^{(1)}(\lambda)=-A^{(1)}(\lambda)-D^{(1)}(\lambda)$ on (38), we have
\begin{eqnarray}
e(\lambda)=-[\prod_{\alpha=1}^M\frac1{a(\lambda-\mu_\alpha)}\prod_{j=1}^n\frac{a(\lambda-k_j)}{a(k_j-\lambda)}
+\prod_{\alpha=1}^M\frac 1{a(\mu_\alpha-\lambda)}\prod_{j=1}^na(\lambda-k_j)]
\end{eqnarray}
and the cancellation of the unwanted terms gives the Bethe ansatz equation
\begin{eqnarray}
\prod_{i=1}^n a(k_i-\mu_\alpha)=\prod_{\alpha\neq \beta}^M\frac{a(\mu_\beta-\mu_\alpha)}{a(\mu_\alpha-\mu_\beta)}.
\end{eqnarray}
To ensure (24) to be an eigenstate of $\tau(\lambda)$, the unwanted terms in (25) and (26) must cancel, i.e.,
\begin{eqnarray}
[-(\Lambda_l)^{b_1\cdots b_n}_{a_1\cdots a_n}+({\bar\Lambda}_l)^{b_1\cdots b_n}_{a_1\cdots a_n}]F^{a_n\cdots a_1}=0.
\end{eqnarray}
This gives
\begin{eqnarray}
a^{N}(k_l+i\frac g2)e^{ik_lL}=\prod_{\alpha=1}^M a(k_l-\mu_\alpha).
\end{eqnarray}
(For the detailed derivation of (30), (31) and (42), we refer the readers to the appendix B of Ref.[12], while the
algebraic structure of the present model is almost the same to that of theirs.) The eigenvalue of $\tau(\lambda)$ reads
\begin{eqnarray}
\nu(\lambda,\{k_j\},\{\mu_\alpha\})=a^N(\lambda+i\frac g2)e^{i\frac\lambda2L}\prod_{j=1}^{N_e}\frac1{a(\lambda-k_j)}e(\lambda)
+e^{-i\frac\lambda2L}\prod_{j=1}^{N_e}\frac 1{a(k_j-\lambda)},
\end{eqnarray}
where $N_e=n$ is the electron number and $M$ is the number of electron with down spins.
Putting $\mu_\alpha=\lambda_\alpha-ig/2$, the Bethe ansatz equations (40) and (42) are reduced to
 
\begin{eqnarray}
e^{ik_jNa}\left(\frac{k_j+\frac i2g}{k_j-\frac i2g}\right)^N=\prod_{\alpha=1}^M\frac
{k_j-\lambda_\alpha+\frac i2g}{k_j-\lambda_\alpha-\frac i2g},\\
\prod_{j=1}^{N_e}\frac{\lambda_\alpha-k_j-\frac i2g}{\lambda_\alpha-k_j+\frac i2g}=-\prod_{\beta=1}^M
\frac{\lambda_\alpha-\lambda_\beta-ig}{\lambda_\alpha-\lambda_\beta+ig},
\end{eqnarray}
\par
and the eigenvalue of $\tau(\lambda)$ can be written clearly as
\begin{eqnarray}
\nu(\lambda,\{k_j\},\{\lambda_\alpha\})=e^{-i\frac\lambda 2L}\prod_{j=1}^{N_e}(1+\frac{ig}{\lambda-k_j})\nonumber\\
-e^{i\frac\lambda2L}\left(\frac{\lambda+i\frac g2}{\lambda-i\frac g2}\right)^N\{\prod_{\alpha=1}^M\frac{\lambda-\lambda_\alpha-
i\frac g2}
{\lambda-\lambda_\alpha+i\frac g2}\prod_{j=1}^{N_e}(1+\frac{ig}{\lambda-k_j})+
\prod_{\alpha=1}^M(1+\frac{ig}{\lambda-\lambda_\alpha+i\frac g2})\}.
\end{eqnarray}
Now we turn to the construction of the model Hamiltonian. For $\lambda\to\infty$ in the upper half complex plane, we have
the following asymptotic expansion
\begin{eqnarray}
\ln[\tau(\lambda)e^{i\frac\lambda2L}]=1+ig\{\frac{C_1}\lambda+\frac{C_2}{\lambda^2}+\frac{C_3}{\lambda^3}+\cdots\}.
\end{eqnarray}
From the commutation relation (19) we know
\begin{eqnarray}
[C_m, C_n]=0, {~~~~}m,n=1,2,3,\cdots.
\end{eqnarray}
In principle, we have more freedoms to choose a Hamiltonian from the conserved
quantities $\{C_n\}$. In this paper, we define the Hamiltonian as
\begin{eqnarray}
H=C_3+igC_2-\frac12g^2C_1.
\end{eqnarray}
For unit boundary condition $T_{n+1}(na,na)=1$, (49) can be derived via Neumann expansion as\cite{11}
\begin{eqnarray}
H_0=\sum_{\sigma}\int\frac{\partial c_\sigma^\dagger(x)}{\partial x}\frac{\partial c_\sigma(x)}{\partial x}
+2g\int c_\uparrow^\dagger(x)c_\downarrow^\dagger(x)c_\downarrow(x)c_\uparrow(x)dx,
\end{eqnarray}
which is nothing but the Hamiltonian of a 1D electron gas with $\delta$-potential interactions.
With the local orbitals, (49) reads
\begin{eqnarray}
H=H_0+H_I,
\end{eqnarray}
where $H_I$ is very complicated, which contains a hybridization term 
$\sum_{n,k}V_{n,k}c_\sigma^\dagger(k)X^n_{0,\sigma}+h.c.$, a correlation term $ \sum_{n,\sigma,\sigma'}U_{\sigma,\sigma'}
c_\sigma^\dagger(na)c_\sigma(na)X_{\sigma',\sigma'}^n$ and other irrelevant terms which often appear in the integrable
impurity models\cite{7}. Comparing (47) and (46), we easily obtain the spectrum of the Hamiltonian
is given by
\begin{eqnarray}
E=\sum_{j=1}^{N_e}k_j^2.
\end{eqnarray}
\section{attractive interaction case}
We discuss first the attractive interaction, i.e., $g<0$ case. 
Without the local orbitals, the ground state is a Fermi sea filled by 
the Cooper-pair-like bound pairs. Now by studying (44) and (45),
we can show how the local levels behave in the ground state. 
Carefully checking the Bethe ansatz equations
we find the  the Cooper pair states described by
\begin{eqnarray}
k_\alpha^\pm=\lambda_\alpha\pm\frac i2g,
\end{eqnarray}
are still possible solutions in the thermodynamic limit $L\to \infty$,
despite the existence of the local orbitals, where $\lambda_\alpha$ are real. To study the stability of these pair
states, we consider a reference state, i.e., all the $N_e$ electrons form $N_e/2$ pairs. In this case, the Bethe ansatz
equations are reduced to
\begin{eqnarray}
e^{2i\lambda_\alpha Na}\left(\frac{\lambda_\alpha-i|g|}{\lambda_\alpha+i|g|}\right)^N=
\prod_{\beta=1}^{N_e/2}\frac{\lambda_\alpha-\lambda_\beta-i|g|}{\lambda_\alpha-
\lambda_\beta+i|g|},
\end{eqnarray}
and the energy of this state is given by
\begin{eqnarray}
E=\sum_{\alpha=1}^{N_e/2}2\lambda_\alpha^2-\frac{N_eg^2}{4}.
\end{eqnarray}
Taking the logarithm of (54) we obtain
\begin{eqnarray}
2\lambda_\alpha+\frac1a\theta(\lambda_\alpha)=\frac{2\pi I_\alpha}L+\sum_{\beta=1}^{N_e/2}\theta(\lambda_\alpha-\lambda_\beta),
\end{eqnarray}
where $\theta(x)=2\tan^{-1}(x/|g|)$ and $I_\alpha$ are integers or half integers depending on the parity of $N-N_e/2$. Notice
that each $I_\alpha$ corresponds a pair state and they must be different from each other due to the exclusion principle.
The minimum state (with lowest energy) is thus described by a sequence of $\{I_\alpha\}=\{-(N_e/2-1)/2,\cdots,(N_e/2-1)/2\}$.
In the thermodynamic limit $N_e/L=n\to finite$, the distribution of $\lambda_\alpha$ can be described by a density function
\begin{eqnarray}
\rho(\lambda_\alpha)=\lim_{L\to\infty}\frac1{(\lambda_{\alpha+1}-\lambda_\alpha)L},
\end{eqnarray}
which satisfies
\begin{eqnarray}
\rho(\lambda)=\frac1\pi+\frac 1{a}f(\lambda)-\int_{-\lambda_F}^{\lambda_F}
f(\lambda-\lambda'))\rho(\lambda')\lambda'
\end{eqnarray}
where the cutoff $\lambda_F$ is given by
\begin{eqnarray}
\int_{-\lambda_F}^{\lambda_F}\rho(\lambda)d\lambda=\frac n2,
\end{eqnarray}
and $f(\lambda)=|g|/\pi(\lambda^2+g^2)$. 
Without the local orbitals, the density distribution of $\lambda$ in the ground state takes the form
\begin{eqnarray}
\rho_0(\lambda)=\frac1\pi-\int_{-\lambda_F^0}^{\lambda_F^0}
f(\lambda-\lambda')\rho_0(\lambda')d\lambda'\\
\int_{-\lambda_F^0}^{\lambda_F^0}\rho_0(\lambda)d\lambda=\frac n2.
\end{eqnarray}
Comparing (58) and (60) we can readily read off $\lambda_F<\lambda_F^0$. That means the effective Fermi energy is 
reduced by the hybridization relative to that of the homogeneous system, a typical heavy fermion 
behavior\cite{2}.
Now we consider the density of states at the Fermi surface. The energy density of the minimum state (relative to the chemical
potential) can be written as
\begin{eqnarray}
E/L=\int_{-\lambda_F}^{\lambda_F}(2\lambda^2-\frac{g^2}2-2\mu)\rho(\lambda)d\lambda,
\end{eqnarray}
where $\mu$ is the chemical potential. Substituting (58) into (62) we readily get
\begin{eqnarray}
E/L=\int_{-\lambda_F}^{\lambda_F}[\frac1\pi+\frac 1af(\lambda)]\epsilon(\lambda)d\lambda,
\end{eqnarray}
where $\epsilon(\lambda)$ is the dressed energy\cite{13} of the Cooper pairs
\begin{eqnarray}
\epsilon(\lambda)=2\lambda^2-\frac{g^2}2-2\mu-\int_{-\lambda_F}^{\lambda_F} f(\lambda-\lambda')\epsilon(\lambda')d\lambda'.
\end{eqnarray}
Consider a particle-hole excitation relative to the minimum state. The density of the
hole and that of the excited particle can be expressed as
\begin{eqnarray}
\rho_h(\lambda)=\frac1L\delta(\lambda-\lambda_h),{~~~~}\rho_p(\lambda)=\frac1L\delta(\lambda-\lambda_p),
\end{eqnarray}
where $\lambda_h$ and $\lambda_p$ are the centers of the hole and the particle, respectively. In addition, such an excitation
induces the back flow of the Fermi sea, i.e., $\delta\rho(\lambda)$. From the BAE we have
\begin{eqnarray}
\rho(\lambda)+\delta\rho(\lambda)=\frac1\pi+\frac1a f(\lambda)+\rho_h(\lambda)-\rho_p(\lambda)\nonumber\\
-\int f(\lambda-\lambda')[\rho(\lambda')
+\delta\rho(\lambda')+\rho_p(\lambda')-\rho_h(\lambda')]d\lambda'.
\end{eqnarray}
With (58) we have
\begin{eqnarray}
\delta\rho(\lambda)=\rho_h(\lambda)-\rho_p(\lambda)+\frac1L[f(\lambda-\lambda_p)-f(\lambda-\lambda_h)]-\int_{-\lambda_F}^{\lambda_F}
f(\lambda-\lambda')\delta\rho(\lambda')d\lambda'.
\end{eqnarray}
The excitation energy reads
\begin{eqnarray}
\delta E=L\int_{-\lambda_F}^{\lambda_F}2\lambda^2\delta\rho(\lambda)d\lambda-2\lambda^2_h+2\lambda^2_p.
\end{eqnarray}
Substituting (67) into (68) we readily obtain that
\begin{eqnarray}
\delta E=\epsilon(\lambda_p)-\epsilon(\lambda_h).
\end{eqnarray}
Therefore, $\epsilon(\lambda)$ can be treated as the quasi-particle energy of the elementary excitations. The excitation
of breaking a Cooper pair can be treated in a similar process. Such an excitation can be described by a $\lambda$-hole in
the Fermi sea
and two real $k$-modes $k_1, k_2$ above the Fermi level. In this case, the excitation energy is
\begin{eqnarray}
\delta E=-\epsilon(\lambda_h)-2\mu+k_1^2+k_2^2+\frac{g^2}2.
\end{eqnarray}
Notice the dressed energy has the properties $\epsilon(\pm \lambda_F)=0$, $\epsilon(\lambda)<0$ for $|\lambda|<\lambda_F$
and $\epsilon(\lambda)>0$ for $|\lambda|>\lambda_F$. Concerning the excitation near the Fermi surface, i.e., 
$\lambda_h\to\lambda_F$, $k_1^2, k_2^2\to \mu$, from (70) we readily obtain that there is a finite gap $g^2/2$ to break 
a Cooper-pair at the Fermi surface. The energy gap seems not changed via the local orbitals. The minimum state we introduced
is thus the absolute ground state of the system for a given $N_e$.
From the BAE (56) we can see  the ``quasiparticle" (Cooper pair) momenta can be defined as
 $p(\lambda_\alpha)=2\pi I_\alpha/L$, then in our
case $p'(\lambda)=2\pi\rho(\lambda)$. The density of states at the Fermi surface is thus
\begin{eqnarray}
N(\lambda_F)=\frac 1\pi\frac{dp(\lambda)}{d\epsilon(\lambda)}|_{\lambda=\lambda_F}=\frac{2\rho(\lambda_F)}
{\epsilon'(\lambda_F)}\equiv \frac1{\pi v},
\end{eqnarray}
where $v$ is the sound velocity\cite{9,13,14}. Since $\rho(\lambda_F)>\rho_0(\lambda_F^0)$ and $\epsilon'(\lambda_F)$
is an increasing function of $\lambda_F$ as can be shown in (64), we deduce that the density of states is enlarged
by the local orbitals. It is not very strange because the Fermi sphere is compressed.
\par
At present, we can see that the local orbitals can not destroy the Cooper pair state completely. However,
it is still not clear whether the local orbitals weaken the pair correlation or enhance it. 
To answer this question,
let us consider the stiffness constant $K$ which measures the non-universal exponents of a variety of
correlation functions in 1D\cite{14}. For the integrable systems, $K=Z^2(\lambda)$ and the dressed charge
function $Z(\lambda)$\cite{13} in our case satisfies
\begin{eqnarray}
Z(\lambda)=1-\int_{-\lambda_F}^{\lambda_F} f(\lambda-\lambda')Z(\lambda')d\lambda'.
\end{eqnarray}
Notice the local orbitals do not change the form of $Z(\lambda)$ but the value of the cutoff $\lambda_F$.
Easily we can show $dZ(\lambda)/d\lambda_F<0$. That means $Z(\lambda)$ is a monotonically decreasing
function of $\lambda_F$. The stiffness constant $K$ is therefore also a monotonically decreasing function
of $\Lambda$. As stated in earlier publications\cite{15}, the gapless 1D quantum system is conformally
invariant at zero temperature and the non-universal exponents of the correlation functions
can be derived from the finite size corrections of the energy spectrum\cite{16,13}: $\delta E=2\pi vx_\nu/L$,
where $x_\nu$ is the scaling dimension (one half the the critical exponent) of the relevant operator.
 In our case, the spin
excitations have a finite gap while the charge ones are gapless. Therefore, the charge sector is conformally
invariant at zero temperature and the asymptotic long-distance superconducting correlators
can be derived from the finite size correction of the ground state energy\cite{13,14,16}.
Notice the pair operator $c_\downarrow^\dagger(x)c_\uparrow^\dagger(x)$ induces a pair number change by one. The
energy change induced by this operator can be calculated by following the standard method introduced in Ref.[13] as
\begin{eqnarray}
\delta E=E_{\frac{N_e}2+1}^g-E_{\frac{N_e}2}^g-2\mu=4\frac{\partial \mu}{\partial N_e} =\frac{\pi v}{2KL},
\end{eqnarray}
where $E_n^g$ is the ground-state energy with $n$ Cooper pairs.
Hence the pair correlator reads
\begin{eqnarray}
<c_{\uparrow}(x)c_{\downarrow}(x),c_\downarrow^\dagger(0)c_\uparrow^\dagger(0)>\sim x^{-\theta}\\
\theta=\frac 1{2K}.\nonumber
\end{eqnarray}
Since $\theta<\theta_0$ (where $\theta_0$ is the corresponding exponent of the homogeneous system), we conclude
that the local orbitals enhance the superconducting correlations. To see it clearly, let us consider the tunneling
of the Cooper pairs through an impurity. The leading tunneling current through the impurity is
\begin{eqnarray}
J(x,y)\sim -i[c_\downarrow^\dagger(x)c_\uparrow^\dagger(x) c_\uparrow(y)c_\downarrow(y)-h.c.], {~~~~~}x\sim0^-,{~~}y\sim0^+,
\end{eqnarray}
where we have put the impurity at the origin. 
From the boundary conformal field theory\cite{17} we have the time correlator of $J$ 
\begin{eqnarray}
[J(t), J(0)]\sim t^{-4\theta}.
\end{eqnarray}
The tunneling conductance can be derived from the Kubo's formula\cite{17} as
\begin{eqnarray}
\sigma(T)\sim T^{2K^{-1}-2}.
\end{eqnarray}
That means the local orbitals enhance the low-temperature tunneling conductance, which provides another evidence of the
enhancement of the superconducting correlation at low temperatures. The Drude weight $D$, which measures
the persistent current in 1D\cite{18}, can also be derived exactly for the present model. It reads $D=Kv$. From
a similar discussion, we can deduce $D$ is reduced by the local orbitals. That means the local orbitals depress
the persistent current.
\section{repulsive interaction case}
Now we turn to the repulsive case ($g>0$). In this case, the Hamiltonian defined in (49) is 
non-Hermitian. We remark
the study on non-Hermitian systems has drawn considerable attention recently for the applications in
a variety of physical situations such as delocalization in the disordered systems\cite{19}, quantum 
chaos\cite{20},
population biology in random media with convection \cite{21} and metal-insulator transitions derived by an 
imaginary vector potential\cite{22}. The spectrum of a
non-Hermitian Hamiltonian generally falls in the complex plane\cite{23}. In our case, from the Bethe ansatz equation
(44) we can see the solutions of $k_j$ are classified into two types: (i)$|\phi(k_j)|=1$; (ii)$|\phi(k_j)|\neq 1$,
where
\begin{eqnarray}
\phi(k)=e^{ika}\frac{k+\frac i2g}{k-\frac i2g}.
\end{eqnarray}
For the Hermitian Hamiltonian, case (i) represents the real modes and case (ii) denotes the string
solutions (Cooper pairs in the present model). Both of these two types of modes give out  real eigenvalues
of the Hamiltonian. When $g>0$ in our case, the Hamiltonian contains a non-Hermitian hybridization term and
a Hermitian term which is nothing but the Hamiltonian of the repulsive $\delta$-potential Fermi gas\cite{11}.
Even for $|\phi(k)|=1$, the Bethe ansatz equations have complex solutions. A typical solution is the imaginary
mode
$k=i\kappa$ with $\kappa a=\ln[|\kappa+g/2|/|\kappa-g/2|$ in the thermodynamic limit. An interesting 
feature is that the Bethe ansatz equations
allow Cooper pair solutions. For $|\phi(k_j)|>1$, the left hand side of (44) is divergent in the thermodynamic limit
while for $|\phi(k_j)|<1$ it tends to zero. Therefore, the pair solutions  (53) are allowed. However, there
is a constraint for the real part of the pair solutions. The condition $|\phi(k)|\neq 1$ hints
\begin{eqnarray}
e^{-\frac{g}2a}|1+\frac{ig}\lambda|>1,
\end{eqnarray}
which gives the critical value of the real part of the pair solutions
\begin{eqnarray}
\lambda_c=\frac{g}{\sqrt{e^{ga}-1}}.
\end{eqnarray}
The Cooper pair solution can exist only in the region $-\lambda_c<\lambda<\lambda_c$. For the complex solutions
in case(i), $|Im k|>|g|/2$ when $|Re k|<\lambda_c$. Hence the ground state  consists of  these modes and there is
no Cooper pair in the ground state. Even so, we can see the local orbitals do cause some fluctuation towards
the formation of Cooper pairs, since without these orbitals, the system does not have any bound state
of electrons.
\section{conclusion}
In conclusion, we propose an integrable hybridization model for a 1D correlated electron system. We
remark though the local orbitals are introduced periodically, the problem is still at the level of
single impurity since there is no correlation among the local states. In fact, the Bethe ansatz equations
(44) and (45) do not depend on the distribution of the local orbitals in the real space. It remains still
a shortcoming of the integrable models with many impurities since the impurities introduced in such a way
are completely transparent to the host electrons and only the forward scatterings are included. Even so,
a finite concentration of impurities enhances the superconducting correlation in our model. It seems that our
result contradicts to that of Ref.[8]. The difference comes from (i) in our model, the binding
energy is independent of the momenta
of the pairs while in the attractive Hubbard model, it depends on the
pair momenta ($\Lambda$ in [8]); (ii) there is no $E_{imp}$ in the formal expression of the eigenenergy
(52), though the local orbitals change the distribution of $\{k_j\}$; (iii) in our model, the density 
of electrons is not changed by the
local orbitals. This corresponds to the hybridization configurations of $f^0
\rightleftharpoons f^{1}$ for the
local orbitals with a configuration $f^0$ in the atomic limit.
 We remark impurities with higher spin and finite momenta can also be introduced in our model
with a similar procedure. In this case, the spin momenta are split into two classes. One associated with
the pair states (53) and the other represents the dynamics of the rest spin degrees of freedom of 
the local orbitals. The latter excitations are always gapless as in the spin chains and the spin excitations
breaking a pair are still gapful. 
\par
One of the authors (YW) acknowledges the financial supports of Alexander von Humboldt-Stiftung and
China National Foundation for Natural Science.

\end{document}